\documentclass{article}
\usepackage{amsmath, amssymb, amsfonts}
\title{Black hole interior in Painleve-Gullstrand coordinates } 
\author{Hristu Culetu, \\Ovidius University, Dept.of Physics, \\B-dul Mamaia 124, 900527 Constanta, Romania, \\e-mail : hculetu@yahoo.com}

\begin{document}
\numberwithin{equation}{section}
\pagenumbering{arabic}
\maketitle
\newcommand{\fv}{\boldsymbol{f}}
\newcommand{\tv}{\boldsymbol{t}}
\newcommand{\gv}{\boldsymbol{g}}
\newcommand{\OV}{\boldsymbol{O}}
\newcommand{\wv}{\boldsymbol{w}}
\newcommand{\WV}{\boldsymbol{W}}
\newcommand{\NV}{\boldsymbol{N}}
\newcommand{\hv}{\boldsymbol{h}}
\newcommand{\yv}{\boldsymbol{y}}
\newcommand{\RE}{\textrm{Re}}
\newcommand{\IM}{\textrm{Im}}
\newcommand{\rot}{\textrm{rot}}
\newcommand{\dv}{\boldsymbol{d}}
\newcommand{\grad}{\textrm{grad}}
\newcommand{\Tr}{\textrm{Tr}}
\newcommand{\ua}{\uparrow}
\newcommand{\da}{\downarrow}
\newcommand{\ct}{\textrm{const}}
\newcommand{\xv}{\boldsymbol{x}}
\newcommand{\mv}{\boldsymbol{m}}
\newcommand{\rv}{\boldsymbol{r}}
\newcommand{\kv}{\boldsymbol{k}}
\newcommand{\VE}{\boldsymbol{V}}
\newcommand{\sv}{\boldsymbol{s}}
\newcommand{\RV}{\boldsymbol{R}}
\newcommand{\pv}{\boldsymbol{p}}
\newcommand{\PV}{\boldsymbol{P}}
\newcommand{\EV}{\boldsymbol{E}}
\newcommand{\DV}{\boldsymbol{D}}
\newcommand{\BV}{\boldsymbol{B}}
\newcommand{\HV}{\boldsymbol{H}}
\newcommand{\MV}{\boldsymbol{M}}
\newcommand{\be}{\begin{equation}}
\newcommand{\ee}{\end{equation}}
\newcommand{\ba}{\begin{eqnarray}}
\newcommand{\ea}{\end{eqnarray}}
\newcommand{\bq}{\begin{eqnarray*}}
\newcommand{\eq}{\end{eqnarray*}}
\newcommand{\pa}{\partial}
\newcommand{\f}{\frac}
\newcommand{\FV}{\boldsymbol{F}}
\newcommand{\ve}{\boldsymbol{v}}
\newcommand{\AV}{\boldsymbol{A}}
\newcommand{\jv}{\boldsymbol{j}}
\newcommand{\LV}{\boldsymbol{L}}
\newcommand{\SV}{\boldsymbol{S}}
\newcommand{\av}{\boldsymbol{a}}
\newcommand{\qv}{\boldsymbol{q}}
\newcommand{\QV}{\boldsymbol{Q}}
\newcommand{\ev}{\boldsymbol{e}}
\newcommand{\uv}{\boldsymbol{u}}
\newcommand{\KV}{\boldsymbol{K}}
\newcommand{\ro}{\boldsymbol{\rho}}
\newcommand{\si}{\boldsymbol{\sigma}}
\newcommand{\thv}{\boldsymbol{\theta}}
\newcommand{\bv}{\boldsymbol{b}}
\newcommand{\JV}{\boldsymbol{J}}
\newcommand{\nv}{\boldsymbol{n}}
\newcommand{\lv}{\boldsymbol{l}}
\newcommand{\om}{\boldsymbol{\omega}}
\newcommand{\Om}{\boldsymbol{\Omega}}
\newcommand{\Piv}{\boldsymbol{\Pi}}
\newcommand{\UV}{\boldsymbol{U}}
\newcommand{\iv}{\boldsymbol{i}}
\newcommand{\nuv}{\boldsymbol{\nu}}
\newcommand{\muv}{\boldsymbol{\mu}}
\newcommand{\lm}{\boldsymbol{\lambda}}
\newcommand{\Lm}{\boldsymbol{\Lambda}}
\newcommand{\opsi}{\overline{\psi}}
\renewcommand{\tan}{\textrm{tg}}
\renewcommand{\cot}{\textrm{ctg}}
\renewcommand{\sinh}{\textrm{sh}}
\renewcommand{\cosh}{\textrm{ch}}
\renewcommand{\tanh}{\textrm{th}}
\renewcommand{\coth}{\textrm{cth}}

\begin{abstract}
A nonstatic Schwarzschild black hole interior solution in Painleve-Gullstrand coordinates is proposed in this paper, by means of a coordinate transformation that changes the spatial coordinate but the timelike one is preserved. The timelike and null geodesic equations are obtained exactly, taking advantage that the spatial z-coordinate is cyclic.
 \end{abstract}

\section{Introduction}
 The Schwarzschild (S) solution of the gravitational field equations has a fundamental importance in the conceptual discussions of General Relativity. The exterior geometry of the S solution was extremelly successful in explaining the phenomenon of the light bending, Mercury's perihelion precession or red-shift effect of the light frequences. However, there are certain ambiguities in the S solution \cite{DLC} - the presence of an event horizon (a one-way membrane), of a singularity (naked or hidden behind the event horizon) or of a signature flip when the black hole (BH) horizon is crossed: the exterior radial coordinate $r_{s}$ and temporal coordinate $t_{s}$ reverse their role inside the BH. Thus, the interior solution becomes a nonstatic geometry \cite{RB}.

Dolan et al. \cite{DLC} studied the interior S solution in detail. They remind that the central singularity located at $r_{s} = 0$ is a spacelike hypersurface and the test particles are not directed towards a priviledged point. They obtained the line element for the interior region without the prejudices inherited from the exterior region. In addition, they found the interior Eddington-Finkelstein line element directly from the interior S solution, investigating the corresponding timelike and null geodesics. Notice that there are another useful coordinates which are regular at the S horizon - the so-called Painleve-Gullstrand coordinates - which were not studied by the above authors inside of the BH.

A through examination of the BH interior was also made by Brehme in \cite{RB}. He observed that the interior universe in the spatial z-direction (which replaces the exterior radial coordinate) is infinite, with $z \in (-\infty , \infty )$. Moreover, the interior world has a finite lifetime $t \in [0, T]$, where the constant $T$ is the equivalent of $2m$ from the exterior world, $m$ being the BH mass. Brehme showed that the exterior source at $r = 0$ is not a point source inside, but an ''instant'' source. The mass $m$ appears only at the moment $t = 0$ and is uniformly distributed along the z-axis, so that the world inside is established by an initial condition rather than by a boundary condition at spatial infinity. 

 The Painleve-Gullstrand (PG) form \cite{PP, AG} of the S geometry has been less studied (see, however \cite{KW, MP, KSH, KK}). Kraus and Wilczek \cite{KW} investigated the radial null geodesics of the S metric in PG coordinates even for $r < 2m$, observing that one meets no obstruction at the horizon $r = 2m$. Martel and Poisson \cite{MP} noticed the striking property that the four-velocity of a geodesic observer in PG coordinates may be written as a gradient of some scalar function. This property is remarkable and it turns out to follow from the equation of motion. The gravitational collapse in PG coordinates is constructed by Kanai et al.\cite{KSH} using a single coordinate patch. They used a generalized form of the PG coordinates where the time coordinate is the proper time of a freely-falling observer. They gave the solution of Einstein's equations in the cases of the collapse from a finite radius as well as from infinity.

 Kassner \cite{KK} remarked that the spatial coordinates are the same in the S and PG coordinates, only the timelike ones are different. However, they run at the same rate for a stationary coordinate observer at $r$. Inside the horizon the continuity of the PG time across $r = 2m$ suggest that it is a more suitable time coordinate than the S time. He also studied the dynamics of a test particle falling radially towards a BH in PG coordinates with a time dependent mass, both for massive and massless particles.
  
	Our purpose in this paper is to look for the PG line-element inside the horizon of the BH, starting from the nonstatic interior S spacetime, even though the interior geometry is usually considered as a continuity of the exterior geometry. We shall adopt the strategy of Dolan et al. \cite{DLC} and analyze the interior S region without the pre-judgement inherited from the exterior region. However, their recipe we apply to the PG form of the S metric (the authors of \cite{DLC} dealt only with the S and Eddington-Finkelstein forms of the interior spacetime)
	
	The paper is organized as follows: in Sec.2 the nonstatic BH metric in PG coordinates is introduced. The coordinate transformation changes the radial coordinate but the timelike one is preserved. The timelike and null geodesics are calculated in Sec.3. A summary and conclusions are given in Sec.4. The geometrical units G = c = 1 will be used throughout the paper, unless otherwise specified.
	
	\section{Black hole interior in PG coordinates}
	Consider the standard representation of the S line-element
	\begin{equation}
  ds^{2} = -(1- \frac{2m}{r_{s}}) dt_{s}^{2} + (1- \frac{2m}{r_{s}})^{-1} dr_{s}^{2} + r_{s}^{2} d \Omega^{2}, 
 \label{2.1}
 \end{equation}
where $r_{s},~t_{s}$ are the S radial and time coordinate, respectively, $m$ is the BH mass and $d \Omega^{2}$ is the squared line-element on the surface of the unit two-sphere. In the interior of the BH ($r_{s} < 2m$), a signature switch takes place so that $t_{s}$ and $r_{s}$ exchange their roles: $t_{s}$ becomes a spacelike coordinate and $r_{s}$ a timelike coordinate. Therefore, the spacetime (2.1) appears now as \cite{DLC, RB, HC}
	\begin{equation}
  ds^{2} = - \frac{1}{\frac{2T}{t} - 1} dt^{2} + (\frac{2T}{t} - 1) dy^{2} + t^{2} d \Omega^{2},~~~~0 < t < 2T 
 \label{2.2}
 \end{equation}
The constant $T$ may be fixed from a direct comparison with the exterior S solution (the matching condition gives $T = m$). The coordinate $y$ takes the role of the radial coordinate from the exterior region. As Brehme \cite{RB} has noticed, the interior world in the y-direction is infinite, with $-\infty < y < \infty$. In addition, the mass $m$ is distributed uniformly along the y-axis. 

To arrive at the PG form of the above line-element, the following coordinate transformation is performed
	\begin{equation}
	z = y - f(t),
 \label{2.3}
 \end{equation}
whence
	\begin{equation}
	dz = dy - (df(t)/dt) dt.
 \label{2.4}
 \end{equation}
When (2.4) is introduced in (2.2), one obtains
	\begin{equation}
  ds^{2} = \left[(\frac{2m}{t} - 1) \left(\frac{df}{dt}\right)^{2} - \frac{1}{\frac{2m}{t} - 1}\right] dt^{2}   + (\frac{2m}{t} - 1) dz^{2} + 
	         2(\frac{2m}{t} - 1) \frac{df}{dt} dz dt + t^{2} d \Omega^{2} .
 \label{2.5}
 \end{equation}
If we now choose the function $f(t)$ such that the term multiplying $dt^{2}$ is -1, we get \footnote{We chose $df/dt > 0$ for to examine the case $\dot{z} < 0$, as we will see later.}
	\begin{equation}
  ds^{2} = -dt^{2} + (\frac{2m}{t} - 1) dz^{2} + 
	         2 \sqrt{2 - \frac{2m}{t}} dz dt + t^{2} d \Omega^{2} ,~~~~m < t < 2m.
 \label{2.6}
 \end{equation}
We propose this geometry as the spacetime in the interior of a SBH in PG coordinates. The metric (2.6) is, of course, Ricci-flat, but the variable $t$ is restricted between $m$ and $2m$ and $-\infty < z < \infty$ \cite{RB}. It is worth noting that the spatial coordinate $z$ is no longer a radial coordinate. Moreover, even though an exterior observer consider a spherically-symmetric spacetime, that is not valid in the interior where the geometry appears to be planar, as points of different $\phi$-coordinate are parallel to one another \cite{DLC}. In addition, the Kretschmann scalar $K$ is given by $K = 48m^{2}/t^{6}$, i.e. a curvature singularity occurs at $t = 0$. However, the singularity is out of the domain of variation of the time variable, that is $t \in (m, 2m)$. Note also that the coefficient of the off-diagonal term in (2.6) has not exactly the same form as for the exterior PG metric. That is due to our choice of the tt-metric coefficient to be -1.

\section{Geodesics}
We study now the interior region not as a continuation of the exterior domain but as a spacetime with its own properties. To start with, we take advantage that the geometry (2.6) does not depend on the spatial coordinate $z$ and we also consider only geodesics along the z-coordinate ($d\theta = d\phi = 0$). We take a look at a freely-falling massive test particle and obtain its equation of motion  from the Lagrangean
\begin{equation}
L = \frac{1}{2} g_{ab}~ \dot{x}^{a} \dot{x}^{b} ,
 \label{3.1}
 \end{equation}
where $a,b = 0, 1, 2, 3$ and $\dot{x}^{a} = dx^{a}/d\tau$, $\tau$ being the proper time. The Euler-Lagrange equations 
\begin{equation}
\frac{\partial L}{\partial x^{a}} - \frac{d}{d\tau} \frac{\partial L}{\partial \dot{x}^{a}}  = 0
 \label{3.2}
 \end{equation}
yield, keeping in mind that $z$ is a cyclic coordinate
\begin{equation}
 \frac{d}{d\tau} \left[(\frac{2m}{t} - 1) \dot{z} + \sqrt{2 - \frac{2m}{t}}~ \dot{t}\right] = 0
 \label{3.3}
 \end{equation}
whence, along the geodesic
\begin{equation}
 (\frac{2m}{t} - 1) \dot{z} + \sqrt{2 - \frac{2m}{t}} ~\dot{t} = const.
 \label{3.4}
 \end{equation}
From (2.6) we have also
\begin{equation}
 \dot{t}^{2} - (\frac{2m}{t} - 1) \dot{z}^{2} - 2 \sqrt{2 - \frac{2m}{t}} ~\dot{t} \dot{z}= 1.
 \label{3.5}
 \end{equation}
We choose $\dot{z} = 0$ (zero initial momentum at $t = m$) and so we get $const. = 0$ in (3.4). Combining (3.4) and (3.5) one obtains
\begin{equation}
 u^{a} \equiv (\dot{t}, \dot{z}, 0, 0) = \left(\sqrt{\frac{2m}{t} - 1}, - \frac{\sqrt{2 - \frac{2m}{t}}}{\sqrt{\frac{2m}{t} - 1}}, 0, 0\right) ,                                                         
 \label{3.6}
 \end{equation}
where $u^{a} = dx^{a}/d\tau$, with $u^{a}u_{a} = -1$. One could easily check that we have, indeed, $a^{b} = u^{a} \nabla_{a} u^{b} = 0$, where $a^{b}$ is the acceleration four-vector. Eq. (3.6) gives us
\begin{equation}
   \frac{dz}{dt} = - \frac{\sqrt{2 - \frac{2m}{t}}}{\frac{2m}{t} - 1} ,                 
 \label{3.7}
 \end{equation}
whence the equation of motion in terms of the coordinate time could emerge. By means of the following change of the time variable, $u = (2m/t) - 1$, we get
\begin{equation}
z(u) = \int \frac{2m\sqrt{1 - u}}{u(1 + u)^{2}} du \equiv 2m (I_{1} - I_{2} - I_{3}),
 \label{3.8}
 \end{equation}
where 
\begin{equation}
I_{1} = \int \frac{\sqrt{1 - u}}{u} du,~~~I_{2} = \int \frac{\sqrt{1 - u}}{1 + u} du, ~~~I_{3} = \int \frac{\sqrt{1 - u}}{(1 + u)^{2}} du .
 \label{3.9}
 \end{equation}
Calculating the above elementary integrals we have, in terms of the physical variable $t$
\begin{equation}
\begin{split}
z(t) = t \sqrt{2 - \frac{2m}{t}} + 2m ln(\frac{2m}{t} - 1) - 4mln(1 + \sqrt{2 - \frac{2m}{t}})\\ +\frac{6m}{\sqrt{2}} ln(\sqrt{2} + \sqrt{2 - \frac{2m}{t}}) + \frac{3m}{\sqrt{2}}~ lnt - \frac{3m}{\sqrt{2}} ln2m .
\end{split}
 \label{3.10}
 \end{equation}
The constant of integration was chosen such that $z(m) = 0$. From (3.7) it is clear that $z(t)$ is a decreasing function of $t$. We have, indeed, $z(t) \rightarrow -\infty$ when $t \rightarrow 2m$, due to the second term in (3.10). 

 Let us consider now the null geodesics in the geometry (2.6). The relation equivalent to (3.5) is now
\begin{equation}
 \dot{t}^{2} - (\frac{2m}{t} - 1) \dot{z}^{2} - 2 \sqrt{2 - \frac{2m}{t}} ~\dot{t} \dot{z}= 0 ,
 \label{3.11}
 \end{equation}
obtained from $ds^{2} = 0$. The overdot stands here for the derivative with respect to the affine parameter along null geodesic, that is $\dot{t} = dt/d\lambda,~ \dot{z} = dz/d\lambda$. The above equation leads to
\begin{equation}
(\frac{2m}{t} - 1) \left(\frac{dz}{dt}\right)^{2} + 2 \sqrt{2 - \frac{2m}{t}} \frac{dz}{dt} - 1 = 0.
 \label{3.12}
 \end{equation}
We distinguish two situations, corresponding to the two roots of (3.12):

(i)
\begin{equation}
 v_{-} \equiv \frac{dz}{dt} = - \frac{\sqrt{2 - \frac{2m}{t}} + 1}{\frac{2m}{t} - 1} < 0,
\label{3.13}
\end{equation}
with $v_{-} = -1$ at the initial time $t = m$ and  $v_{-} \rightarrow -\infty$ when $t \rightarrow 2m$. Similar calculations as those for the timelike geodesics give us
\begin{equation}
\begin{split}
z(t) = t (\sqrt{2 - \frac{2m}{t}} + 1) - 4m~ln(1 + \sqrt{2 - \frac{2m}{t}}) +\frac{6m}{\sqrt{2}}~ ln(\sqrt{2} + \sqrt{2 - \frac{2m}{t}}) \\ + m (\frac{3}{\sqrt{2}} - 2)~ lnt + 4m~ln(2m - t) - \frac{3m}{\sqrt{2}} ln2m - 2m~lnm - m.
\end{split}
 \label{3.14}
 \end{equation}
The constant of integration was selected such that $z(m) = 0$. In addition, one observes that $z(t) \rightarrow -\infty$ when $t \rightarrow 2m$ thanks to the term proportional to $ln(2m - t)$. 

(ii) 
\begin{equation}
 v_{+} \equiv \frac{dz}{dt} =  \frac{1 - \sqrt{2 - \frac{2m}{t}}}{\frac{2m}{t} - 1} > 0,
\label{3.15}
\end{equation}
with $v_{+} = 1$ at $t = m$ and $v_{+} \rightarrow 1/2$ when $t \rightarrow 2m$, so that $v_{+}$ remains finite for any value of $t$ in its domain of variation, contrary to the previous case. The equation of motion of the massless test particle along the z-axis appears now as
\begin{equation}
\begin{split}
z(t) = t (\sqrt{2 - \frac{2m}{t}} - 1) - 4m~ln(1 + \sqrt{2 - \frac{2m}{t}}) +\frac{6m}{\sqrt{2}}~ ln(\sqrt{2} + \sqrt{2 - \frac{2m}{t}}) \\ + m (\frac{3}{\sqrt{2}} - 2)~ lnt  - \frac{3m}{\sqrt{2}} ln2m + 2m~lnm + m,
\end{split}
 \label{3.16}
 \end{equation}
with $z(m) = 0$ as before. In (3.16) the two terms containing $ln(2m - t)$ were cancelled out and, therefore, $z(t)$ is finite and positive at $t = 2m$. The positiveness is assured by the fact that $z(t)$ is here an increasing function of $t~(v_{+} > 0)$.

\section{Conclusions}
We have provided in this paper simple arguments that the Schwarzschild spacetime is nonstatic inside the BH even in PG coordinates, though is generally considered that one may use the same coordinate patch both inside and outside the BH horizon. The reason comes from the fact that, to get the interior line-element, the starting point should be the nonstatic interior geometry in Schwarzschild coordinates, as the authors of \cite{DLC} did for other coordinate systems. Though the expressions for timelike and null geodesics in the spatial z-direction are complicate, they are however exact and have reasonable properties.


\begin{thebibliography} {9}

\bibitem{DLC}
R. Dolan, F. S. N. Lobo and P. Crawford, Found. Phys. 38, 160 (2008), arXiv: gr-qc/0609042.
\bibitem{RB}
R. Brehme, Am. J. Phys. 45, 423 (1977).
\bibitem{PP}
P. Painleve, C. R. Acad. Sci. (Paris) 173, 677 (1921).
\bibitem{AG}
A. Gullstrand, Arkiv. Math. Astron. Fys. 16, 1 (1922).
\bibitem {KW}
P. Kraus and F. Wilczek, arXiv: gr-qc/9406042.
\bibitem{MP}
K. Martel and E. Poisson, Am. J. Phys. 69, 476 (2001), arXiv: gr-qc/0001069.
\bibitem {KSH}
Y. Kanai, M. Suno and A. Hosoya, Progr. Theor. Phys. 125, 1053 (2011), arXiv: 1008.0470 [gr-qc].
\bibitem{KK}
K. Kassner, arXiv: 1801.00272.
\bibitem{HC}
H. Culetu, Cent. Eur. J. Phys. 6, 317 (2008), arXiv: hep-th/0703168.




\end{thebibliography}
\end{document}